\documentclass[trackchanges,twocolumn,svgnames]{aastex701}
\usepackage{graphicx}
\usepackage{subcaption}
\usepackage{hyperref}
\usepackage[normalem]{ulem} 
\usepackage{xcolor}
\usepackage{enumitem}
\usepackage{amsmath}

\hypersetup{breaklinks=true}
\begin{document}

\title{What ZTF Saw Where Rubin Looked: Anomaly Hunting in DR23}

\correspondingauthor{Maria Pruzhinskaya}
\email{pruzhinskaya@gmail.com}

\author[orcid=0000-0001-7178-0823,gname=Maria, sname='Pruzhinskaya']{Maria V. Pruzhinskaya} 
\affiliation{Sternberg astronomical institute, Lomonosov Moscow State University, Universitetsky pr. 13, Moscow, 119234, Russia}
\email{pruzhinskaya@gmail.com}

\author[orcid=0009-0002-0835-7998, gname=Anastasia, sname='Lavrukhina']{Anastasia D. Lavrukhina}
\affiliation{Sternberg astronomical institute, Lomonosov Moscow State University, Universitetsky pr. 13, Moscow, 119234, Russia}
\email{lavrukhina.ad@gmail.com}

\author[orcid=0009-0003-2133-6144,gname=Timofey, sname='Semenikhin']{Timofey A. Semenikhin} 
\affiliation{Lomonosov Moscow State University, Faculty of Physics, Leninskie Gory, 1-2, Moscow, 119991, Russia}
\affiliation{Sternberg astronomical institute, Lomonosov Moscow State University, Universitetsky pr. 13, Moscow, 119234, Russia}
\email{ofmafowo@gmail.com}

\author[0000-0003-3554-1037]{Alina A. Volnova}
\affiliation{Space Research Institute of the Russian Academy of Sciences (IKI), 84/32 Profsoyuznaya Street, Moscow, 117997, Russia}
\email{alinusss@gmail.com}

\author[0000-0002-6423-1348]{Sreevarsha Sreejith}
\affiliation{Physics Department, University of Surrey, Stag Hill campus, Guildford, GU2 7XH, UK.}
\email{sreevarshasree@gmail.com}

\author[orcid=0000-0001-9388-691X,gname=Vadim, sname='Krushinsky']{Vadim V.~Krushinsky} 
\affiliation{Laboratory of Astrochemical Research, Ural Federal University, Ekaterinburg, ul. Mira d. 19, Yekaterinburg, 620002, Russia}
\email{krussh@gmail.com}

\author[orcid=0000-0001-6728-1423,gname=Emmanuel, sname='Gangler']{Emmanuel Gangler}
\affiliation{Universit\'e Clermont Auvergne, CNRS/IN2P3, LPC, F-63000 Clermont-Ferrand, France}
\email{Emmanuel.Gangler@clermont.in2p3.fr}

\author[0000-0002-0406-076X]{Emille E. O. Ishida}
\affiliation{Universit\'e Clermont Auvergne, CNRS/IN2P3, LPC, F-63000 Clermont-Ferrand, France}
\email{emille.ishida@clermont.in2p3.fr}

\author[0000-0002-5193-9806]{Matwey V. Kornilov}
\affiliation{Sternberg astronomical institute, Lomonosov Moscow State University, Universitetsky pr. 13, Moscow, 119234, Russia}
\affiliation{National Research University Higher School of Economics, 21/4 Staraya Basmannaya Ulitsa, Moscow, 105066, Russia}
\email{kornilov@physics.msu.ru}

\author[0000-0001-7179-7406]{Konstantin L. Malanchev}
\affiliation{The McWilliams Center for Cosmology \& Astrophysics, Department of Physics, Carnegie Mellon University,
Pittsburgh, PA 15213, USA}
\email{kmalanch@andrew.cmu.edu}

\collaboration{all}{The SNAD team}

\begin{abstract}

We present results from the SNAD VIII Workshop, during which we conducted the first systematic anomaly search in the ZTF fields also observed by LSSTComCam during Rubin Scientific Pipeline commissioning. Using the PineForest active anomaly detection algorithm, we analysed four selected fields (two galactic and two extragalactic) and visually inspected 400 candidates. As a result, we discovered six previously uncatalogued variable stars, including RS~CVn, BY Draconis, ellipsoidal, and solar-type variables, and refined classifications and periods for six known objects. These results demonstrate the effectiveness of the SNAD anomaly detection pipeline and provide a preview of the discovery potential in the upcoming LSST data.
\end{abstract}


\section{Introduction}

The advent of large astronomical data sets, which took place in the last few decades, changed significantly how researchers interact with the data. In all areas of observational astronomy, communities were obliged to adapt to unprecedented volumes of photometric observations. It is safe to say that one of the areas most impacted by this new paradigm was the discovery of new, and sometimes unpredicted, astrophysical phenomena. Serendipitous discovery has became more unlikely  with each new survey, which boosted the use of automatic machine learning techniques. The situation is even more drastic when we take into account expected data of the Legacy Survey of Space and time from the Vera C. Rubin Observatory~\citep{2009arXiv0912.0201L}.

In order to enable systematic search for unpredicted, scientifically interesting, phenomena in large data sets, the SNAD team\footnote{\url{https://snad.space/}} \citep{volnova2024} has been developing a series user-friendly anomaly detection (AD) tools  \citep{2023PASP..135b4503M, 2025A&C....5200960K} whose goal is to enable seamless interaction between the expert and the machine, and thus optimize the search for novelty in astronomical data. The pipeline has already proven to be successful in scrutinizing data from the Zwicky Transient Facility\footnote{\url{https://www.ztf.caltech.edu/}} (ZTF; \citealt{Bellm_2019}) under a series of different circumstances, including searches for supernovae \citep{2022NewA...9601846A, 2023A&A...672A.111P} and stellar flares \citep{10.1093/mnras/stae2031}, as well as the identification of artifacts in astronomical data \citep{2025A&C....5100919S, 2025arXiv250408053S}. We start now to redirect this experience towards a new paradigm.

In this work, we present results from the first systematic anomaly search in fields observed by both, ZTF and the LSST commissioning camera (LSSTComCam, \citealt{guy_2025_15558559}).  
We applied the SNAD pipeline to ZTF Data Release 23 (DR23) light curve data, limited to regions targeted during the on-sky campaign of the LSSTComCam dedicated to the commissioning of its Science Pipelines. Recent reports already presented result from variability search within LSST proprietary data covering the same area \citep{malanchev2025, carlin2025, choi2025}, however, to our knowledge, a similar AD search was never specifically targeted to these fields.

\begin{figure*}[h!]
\begin{center}
\includegraphics[width=\textwidth]{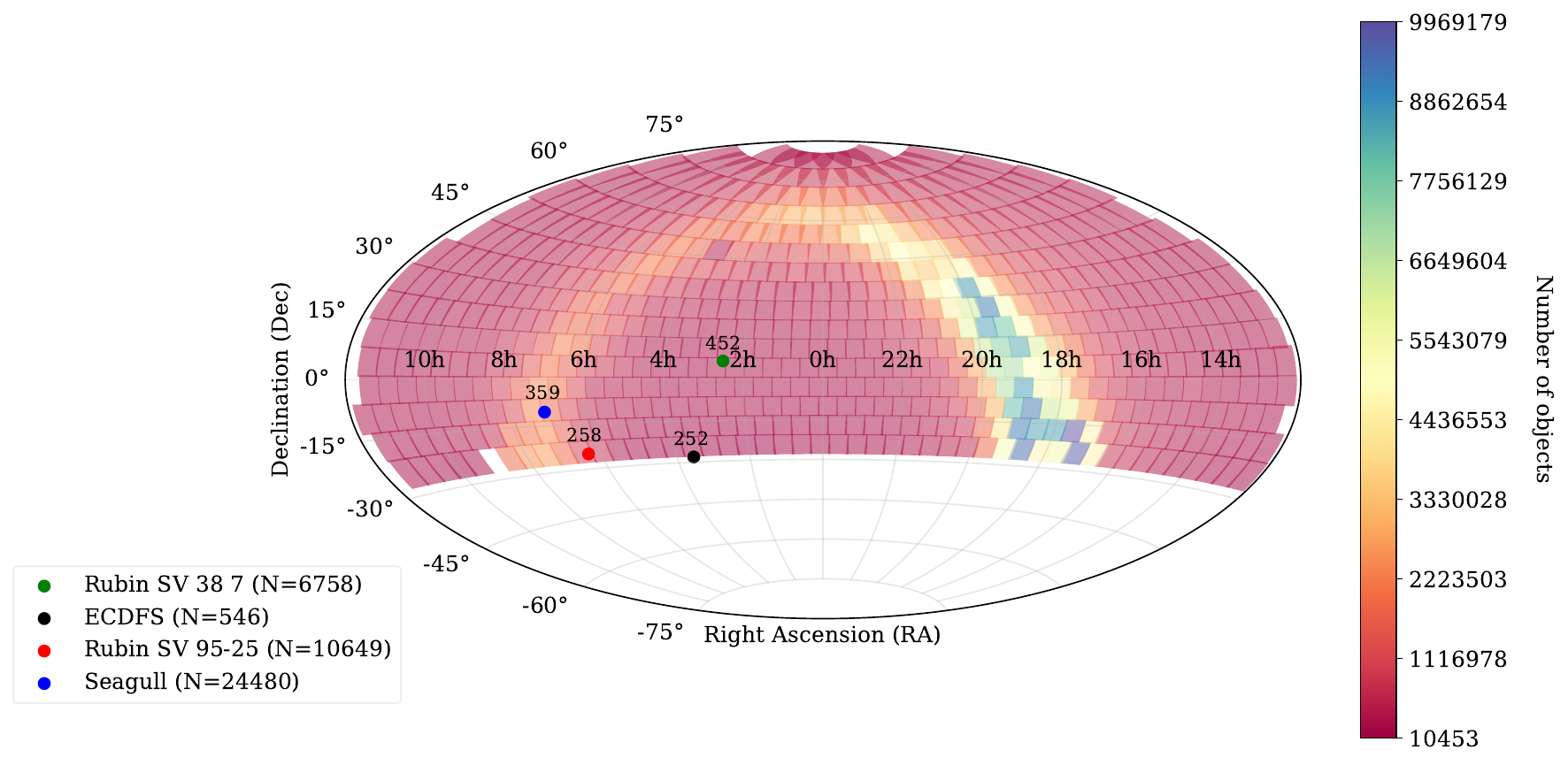}
\caption{Spatial footprint of the four LSSTComCam target fields (circles) overlaid on ZTF sky coverage. Above each circle, the FIELDID of the corresponding overlapping ZTF field is shown. In the legend, the number in parentheses indicates how many ZTF objects fall within the respective LSSTComCam target field. The background grid outlines the ZTF observed fields. Each is colour-coded according to the number of ZTF objects in it.  \label{fig:map}}
\end{center}
\end{figure*}

LSSTComCam is the Rubin Observatory's engineering camera, used for testing and validating the observatory systems prior to the installation of the full LSST Camera \citep{2024SPIE13096E..1SR,lange:hal-04775095}. Its focal plane consists of a single raft with a $3 \times 3$ mosaic of $4\mathrm{K} \times 4\mathrm{K}$ ITL science sensors, totaling 144 million pixels. It shares the same plate scale as the LSSTCam (0.2 arcsec/pixel) and covers a field of view of $40 \times 40$ arcmin. Observations with LSSTComCam were conducted at the end of 2024. For the Science Pipelines commissioning campaign, seven target fields were selected\footnote{\url{https://sitcomtn-149.lsst.io/}}. These fields cover a total area of approximately 15 square degrees. The photometric data from 1791 LSSTComCam exposures formed the Rubin Data Preview 1 (DP1) dataset~\citep{RTN-095}. Among the seven LSSTComCam target fields, four overlap with the ZTF footprint (see Table~\ref{tab:fields}). Figure~\ref{fig:map} shows the spatial footprint of these selected fields overlaid on ZTF coverage.

In what follows, data selection is described in Section \ref{sec:data}; the algorithm used for anomaly search is given in Section \ref{sec:methods}; a summary of our findings is shown in Section \ref{sec:results} and our conclusions are discussed in Section \ref{sec:conclusions}.

\begin{table*}[h!]
\centering
\caption{Coordinates, radii and names of selected LSSTComCam fields. \textbf{N} is number of ZTF objects within the footprint of these fields.}
\begin{center}
\begin{tabular}{llllll}
\hline
\textbf{Field Code} & \textbf{Field Name} & \textbf{RA (deg)} & \textbf{Dec (deg)} & \textbf{Radius (deg)} & \textbf{N} \\
\hline
Rubin SV 38 7     & Low Ecliptic Latitude Field       & 37.86  & $+6.98$   & 0.7 & 6758  \\
ECDFS             & Extended Chandra Deep Field South & 53.13  & $-28.10$  & 0.6 & 546    \\
Rubin SV 95 $-25$ & Low Galactic Latitude Field       & 95.00  & $-25.00$  & 0.5 & 10649 \\
Seagull           & Seagull Nebula                    & 106.23 & $-10.51$  & 0.5 & 24480 \\
\hline
\end{tabular}
\end{center}
\label{tab:fields}
\end{table*}

\section{Data}
\label{sec:data}

We used data from the latest ZTF data release (DR23),  covering the time range of $58178 \leq \mathrm{MJD} \leq 60125$, which corresponds to March 2018 through June 2023~\citep{Bellm_2019}. From this dataset, we selected only $zr$-band light curves that satisfy the quality flag criterion \texttt{catflags = 0}, holding at least 100 detections and overlapping with Rubin commissionning fields.

Feature extraction from ZTF light curves was performed using the \texttt{light-curve}\footnote{\url{https://github.com/light-curve/light-curve-python}} package, a time-series feature extraction tool developed as part of the SNAD pipeline~\citep{2021MNRAS.502.5147M,lcpaper}. The resulting dataset contains 47 statistical features per object; the full list of features and the feature-extracted dataset are available on-line\footnote{\url{https://snad.space/features/}}.

LSSTComCam covers a significantly smaller area than a single ZTF observation field of view of 47 square degrees. Therefore, from the ZTF fields overlapping with a corresponding LSSTComCam target field, we extracted objects likely to also be observed by Rubin. However, the employed LSSTComCam observation strategy involved a complex pattern of random translational and rotational dithers around each field's nominal pointing center, making the exact footprint non-trivial to define.

\begin{figure*}[h!]
\begin{center}
\includegraphics[scale=0.3,angle=0]{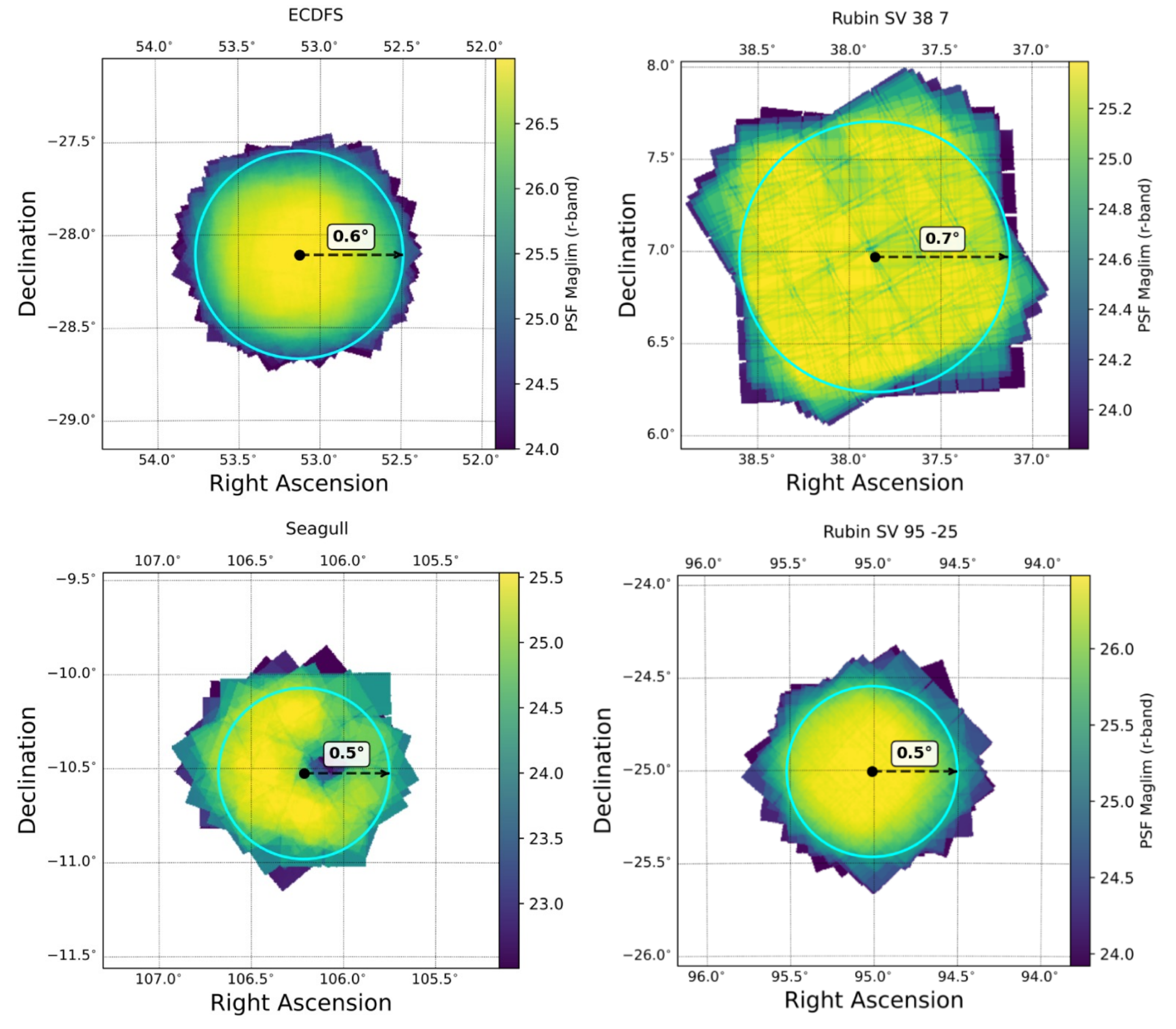}
\caption{Cumulative imaging depth expressed in terms of the $S/N = 5$ limiting magnitude for
unresolved sources for four LSSTComCam fields. We additionally specify a selected radius and corresponding area for each field. \label{fig:fields}}
\end{center}
\end{figure*}

Since LSSTComCam coverage maps are not publicly available in a machine-readable form, we used Fig. 3 of \cite{guy_2025_15558559} for the region selection.
We applied a percentile-based intensity thresholding to identify dark regions corresponding to observational coverage in this figure, with the percentile threshold manually selected for each image to optimize region detection. A distance transform algorithm was then used to determine the maximum inscribed circle within the primary observation region, and to estimate the radius of the corresponding field (Fig.~\ref{fig:fields}).

The radius of each LSSTComCam target field and the number of ZTF objects within the footprint of these fields is given in Table~\ref{tab:fields}.

\section{Methods}
\label{sec:methods}

We applied the PineForest algorithm, implemented in the \texttt{coniferest} library~\citep{2025A&C....5200960K}, to independently explore potential outliers in each of the four fields.

PineForest is an active AD framework developed by the SNAD team. It is based upon the well-known Isolation Forest algorithm~\citep{isoforest}, which adapts the decision tree method~\citep{breiman2001,loh2014,hunt1966} for outlier detection by exploiting the idea that anomalies tend to be more easily isolated in feature space. PineForest allows the personalization of the structure of the forest by selecting only random trees that agree with the anomaly definition informed by the user.

For each field, we conducted two PineForest sessions with a labelling budget of 50, retraining the model between iterations to incorporate expert feedback and refine the anomaly ranking.

Our main findings are summarized in the next sections.

\section{Results}
\label{sec:results}

We analysed, two extragalactic (Rubin SV 387 and ECDFS) and two  galactic (Rubin SV 95$-$25 and Seagull) fields. All of which contain a certain amount of artifacts and spurious data.

The galactic fields include a wide range of variable stars, such as eclipsing binaries (EB), RR Lyrae, RS CVn systems, long-period variables (LPV), cataclysmic variables (CV), and other known and unknown variable objects. 
Additionally, the Seagull field, located within a nebula, also contains a dozen young stellar object (YSO) candidates, consistent with previous studies showing that YSOs preferentially reside in star-forming regions and molecular clouds (e.g., \citealt{2003ARA&A..41...57L, 2009ApJS..181..321E}).

The extragalactic fields contain several extended sources. In these fields, we primarily found active galactic nuclei (AGN), including quasars (QSO) and one blazar candidate (ZTF OID 452216400005600), one possible supernova (AT~2022zgk), and three known asteroids (Patsy, 2003~SN65, and 2015~GZ65). A small number of stellar variables, such as RS CVn stars and EB, are also present in the extragalactic fields.

Below, we present newly identified variable stars, as well as known variables for which we refined the classification and measured periods. The periods were initially estimated using values provided in the SNAD ZTF Viewer\footnote{\url{https://ztf.snad.space/}} \citep{2023PASP..135b4503M}. It uses a period corresponding to the highest Lomb-Scargle periodogram peak \citep{Lomb1976,Scargle1982}. These preliminary estimates were subsequently fine-tuned using the Variability Search Toolkit (VaST; \citealt{Sokolovsky_2018_VaST}). VaST employs methods developed by \citet{Lafler_1965} and \citet{Deeming_1975}.

\subsection{New variables}

All objects presented below are new variables that are not listed in the Gaia DR3 Variability Catalog~\citep{2023A&A...674A...1G} or other known catalogs of variable stars, including the General Catalog of Variable Stars \citep[GCVS, ][]{gcvs2003,gcvs2017}, the International Variable Star Index \citep[AAVSO VXS, ][]{vsx}, the Large-amplitude variables in Gaia DR2~\citep{2021yCat..36480044M} or the Stellar variability catalog from Gaia DR3 \citep{2023A&A...677A.137M}.

Each of these objects has been added to the SNAD Catalog\footnote{\url{https://snad.space/catalog/}} and assigned a SNAD name.

\subsubsection{SNAD273}\label{star:452216100008087}

The star ZTF OID 452216100008087 (Gaia DR3 19492077713298944) is located at equatorial coordinates  $\alpha = \mathrm{02^h\,30^m\,24.828^s}$, $\delta = +07^\circ\,28'\,8.44''$.

The field is extragalactic, but the object is classified as a star by different surveys (e.g., \citealt{2020ApJS..249....3A,2022MNRAS.512.3662D}). Gaia data suggest that it is a main-sequence star, with a photogeometric distance of $959^{+87}_{-70}$~pc~\citep{2021AJ....161..147B}. 

The ZTF light curve (Fig.~\ref{fig:SNAD273}), as well as Pan-STARRS1 DR2 photometry \citep{2016arXiv161205560C,2020ApJS..251....7F}, reveal low-amplitude, non-periodic variability consistent with solar-type activity.

\subsubsection{SNAD274}\label{star:359205200010128}

The star ZTF OID 359205200010128 (Gaia DR3 3046542589365461376) is located at $\alpha = \mathrm{07^h\,04^m\,34.577^s}$, $\delta = -10^\circ\,18'\,10.01''$. The photogeometric distance is estimated to be $1313^{+151}_{-108}$~pc~\citep{2021AJ....161..147B}. According to Gaia data, the star lies above the main sequence, suggesting it could be a rotating ellipsoidal variable (ELL).

The ZTF folded light curve is shown in Fig.~\ref{fig:SNAD274}. The estimated period is $P = 15.2395$ days.

\subsubsection{SNAD276}\label{star:359205200015003}

The ZTF OID 359205200015003\footnote{This object was also independently found using a different approach, Signature Anomaly Detection~\citep{2025arXiv250616314G}.} (Gaia DR3 3046517850352071040) is located at $\alpha = \mathrm{07^h\,04^m\,38.191^s}$, $\delta = -10^\circ\,31'\,58.69''$. The photogeometric distance is estimated to be $970^{+242}_{-147}$~pc~\citep{2021AJ....161..147B}. 

The ZTF folded light curve is shown in Fig.~\ref{fig:SNAD276}. The estimated period is $P = 6.34625$ days. The presence of outlier points in the light curve may indicate chromospheric activity, which is also typical for RS~CVn-type stars.

\subsubsection{SNAD277}\label{star:359205200019542}

The ZTF OID 359205200019542 (Gaia DR3 3046510501666004224) is located at $\alpha = \mathrm{07^h\,04^m\,37.973^s}$, $\delta = -10^\circ\,43'\,37.78''$. The photogeometric distance is estimated to be $870^{+66}_{-78}$~pc~\citep{2021AJ....161..147B}. Based on Gaia data, the object is consistent with a main-sequence star.

Similar to SNAD273, this object shows signs of solar-type variability (see Fig.~\ref{fig:SNAD277}).

\subsubsection{SNAD278}\label{star:359205300002062}

The ZTD OID 359205300002062 (Gaia DR3 3046437074905139072) is located at $\alpha = \mathrm{07^h\,04^m\,41.726^s}$, $\delta = -10^\circ\,54'\,26.14''$. The photogeometric distance is estimated to be $1373^{+307}_{-206}$~pc~\citep{2021AJ....161..147B}.

The ZTF folded light curve is shown in Fig.~\ref{fig:SNAD278}. Based on its photometric properties, specifically, the presence of a distortion wave, an amplitude of $\sim$0.6~mag in the $zr$-band, and flares, we tentatively classify this object as an RS~CVn-type variable. The estimated period is $P = 2.03916$ days.

\subsubsection{SNAD279}\label{star:359205300031896}

The star ZTF OID 359205300031896 (Gaia DR3 3046431916644875136) is located at $\alpha = \mathrm{07^h\,04^m\,45.173^s}$, $\delta = -10^\circ\,59'\,50.28''$. The photogeometric distance is estimated to be $1246^{+175}_{-143}$~pc~\citep{2021AJ....161..147B}. According to \citealt{2024yCat..36860042H}, it is a member of open cluster Theia~1654.

\begin{figure*}[htbp]
    \centering

    \begin{subfigure}[b]{0.46\textwidth}
        \centering
        \includegraphics[width=\textwidth]{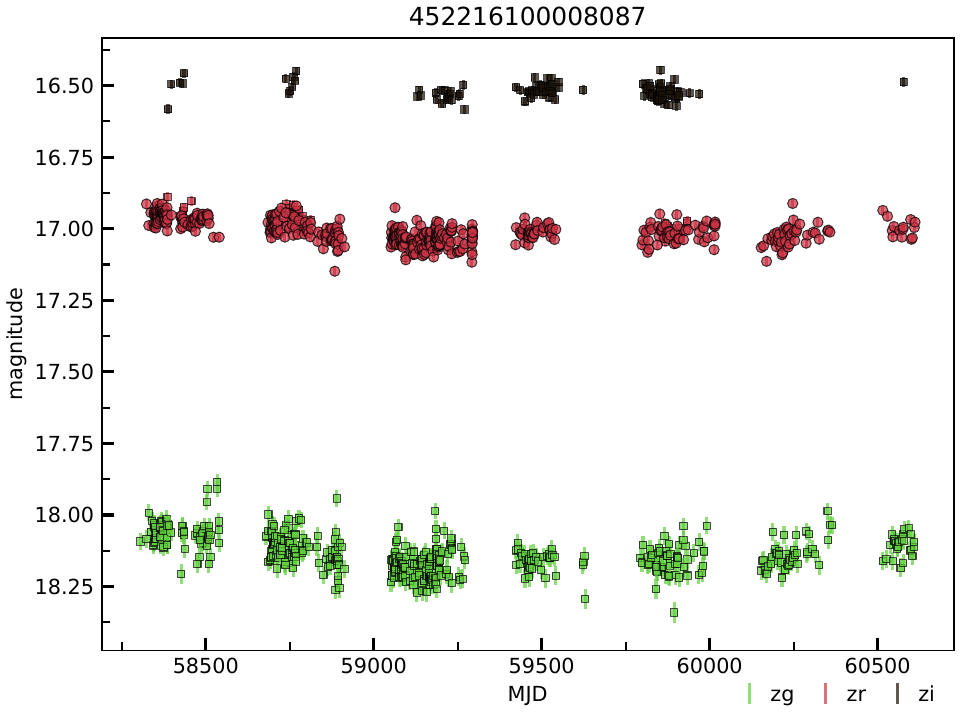}
        \caption{}
        \label{fig:SNAD273}
    \end{subfigure}
    \hfill
    \begin{subfigure}[b]{0.46\textwidth}
        \centering
        \includegraphics[width=\textwidth]{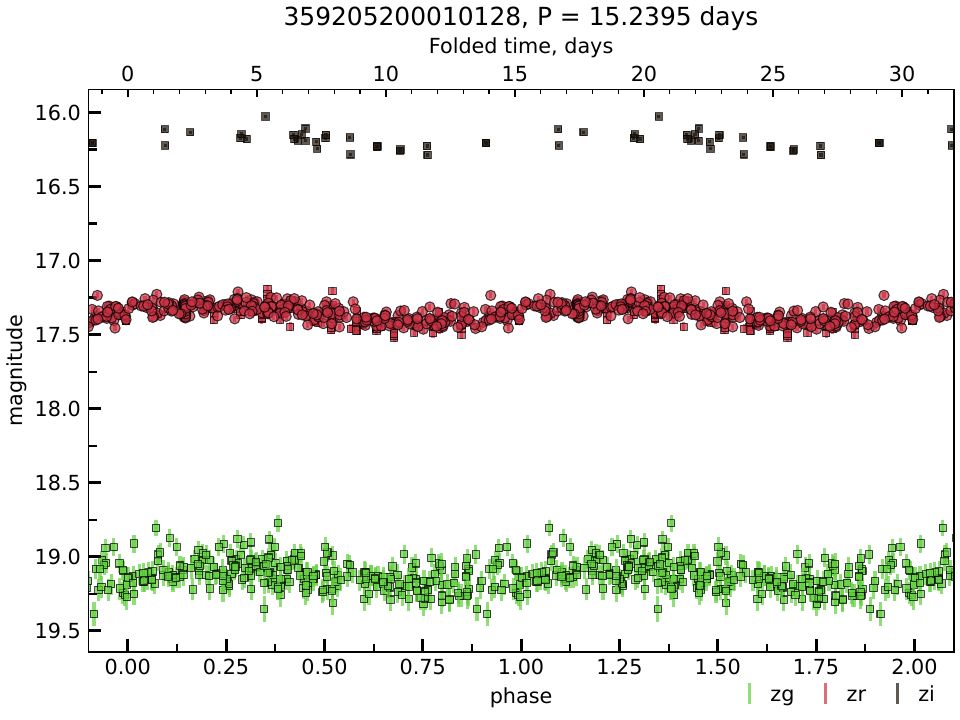}
        \caption{}
        \label{fig:SNAD274}
    \end{subfigure}

    \begin{subfigure}[b]{0.46\textwidth}
        \centering
        \includegraphics[width=\textwidth]{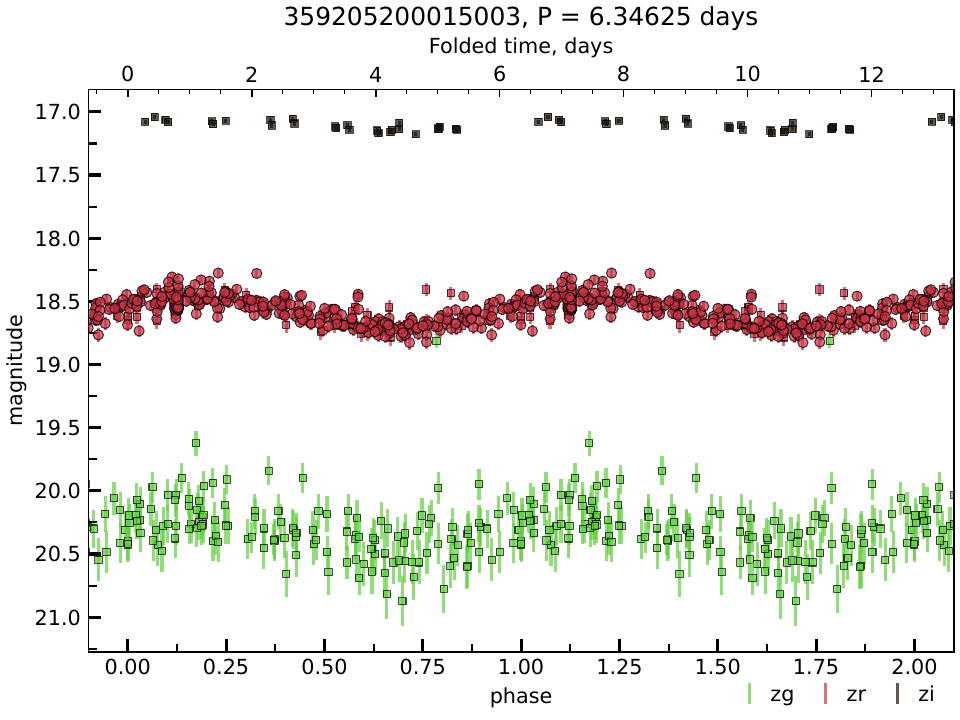}
        \caption{}
        \label{fig:SNAD276}
    \end{subfigure}
    \hfill
    \begin{subfigure}[b]{0.46\textwidth}
        \centering
        \includegraphics[width=\textwidth]{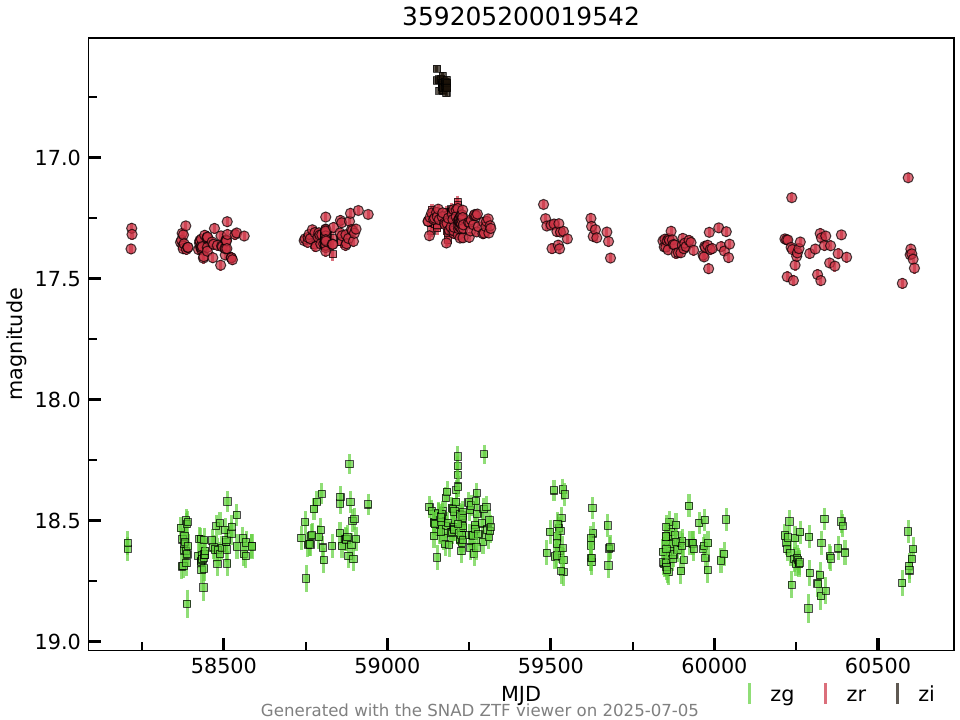}
        \caption{}
        \label{fig:SNAD277}
    \end{subfigure}

    \begin{subfigure}[b]{0.46\textwidth}
        \centering
        \includegraphics[width=\textwidth]{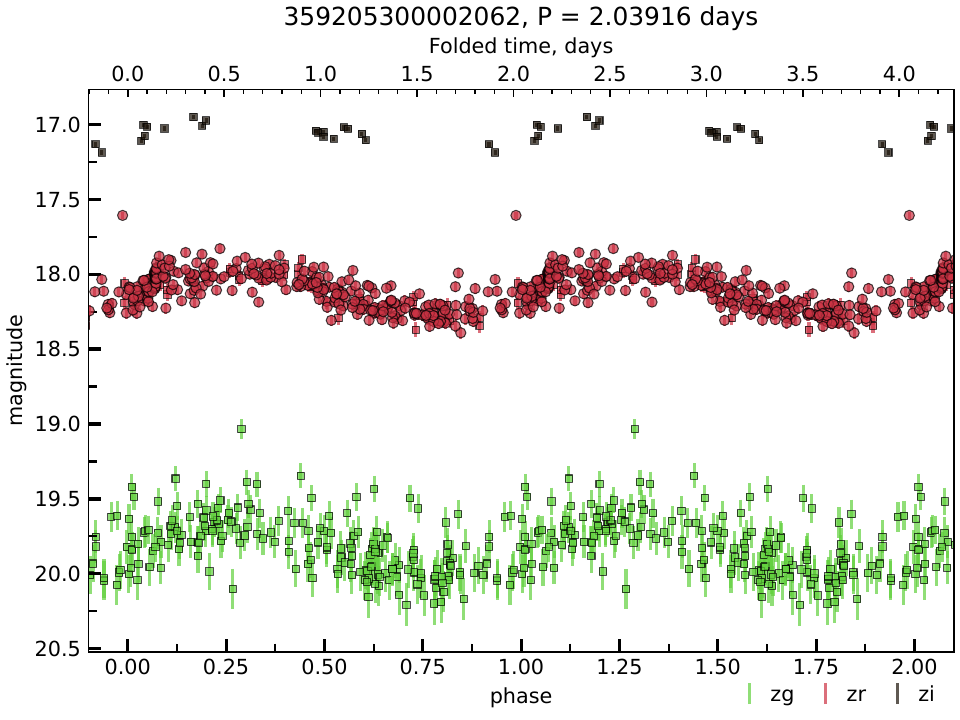}
        \caption{}
        \label{fig:SNAD278}
    \end{subfigure}
    \hfill
    \begin{subfigure}[b]{0.46\textwidth}
        \centering
        \includegraphics[width=\textwidth]{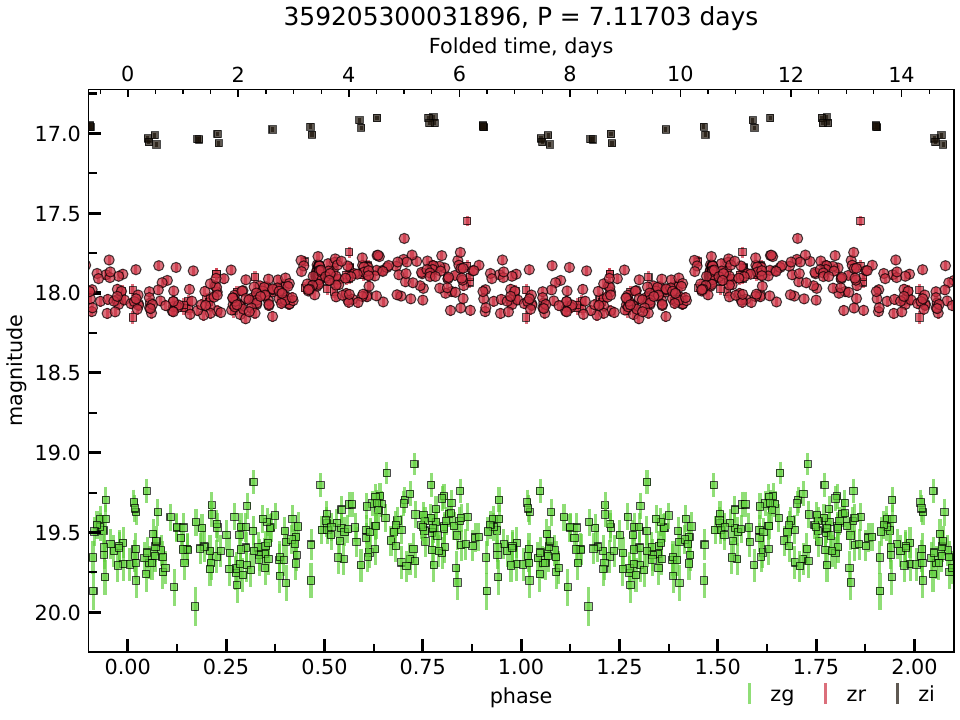}
        \caption{}
        \label{fig:SNAD279}
    \end{subfigure}

    \caption{Light curves of eight new ZTF variable stars discovered in LSSTComCam target fields during the SNAD-VIII Workshop. For periodic variables, folded light curves are shown along with the corresponding periods:
    (\subref{fig:SNAD273}) 452216100008087 (see Section~\ref{star:452216100008087}),
    (\subref{fig:SNAD274}) 359205200010128 (see Section~\ref{star:359205200010128}),
    (\subref{fig:SNAD276}) 359205200015003 (see Section~\ref{star:359205200015003}),
    (\subref{fig:SNAD277}) 359205200019542 (see Section~\ref{star:359205200019542}),
    (\subref{fig:SNAD278}) 359205300002062 (see Section~\ref{star:359205300002062}),
    (\subref{fig:SNAD279}) 359205300031896 (see Section~\ref{star:359205300031896}).}
    \label{fig:combined_new}
\end{figure*}

The ZTF folded light curve is shown in Fig.~\ref{fig:SNAD279}. Based on its photometric properties, we also interpret the observed brightness modulation as likely caused by stellar spots and rotation, BY Draconis type (BY). The estimated period is $P = 7.11703$ days.

\subsection{Classification and determination of light curve parameters of known variables}

In this section, we focus on variable objects that have been previously catalogued as variables, mainly using machine learning techniques.
For these sources, our newly acquired data allowed us to determine their periods and refine or clarify their variability classifications.
This visual re-evaluation provides improved or more precise parameters for known variables, supporting their further study and contributing to the reliability of variable star catalogues.

\subsubsection{452216400002679}\label{star:452216400002679}

The star ZTF OID 452216400002679 (Gaia DR3 19359689641429888) is classified as eclipsing binary (EB) system by \citet{2022yCat.1358....0G}. 
Based on ZTF DR23 data, we revised the period to $1.10166$ days, correcting the inaccurate value of $0.54928$ days previously reported by \citet{2022yCat.1358....0G}. The updated folded light curve is shown in Fig.~\ref{fig:452216400002679}.

\subsubsection{258207200007316}\label{star:258207200007316}
The star ZTF OID 258207200007316 (Gaia DR3 2912319917858743808) is classified as an RS~CVn-type variable by~\citet{2022yCat.1358....0G}, although no period is provided there.

According to Gaia DR3 and EDR3 data, this object is a late G- to early K-type main-sequence star, with an effective temperature of $T_{\mathrm{eff}} = 5212^{+23}_{-21}$~K, surface gravity $\log g = 4.11 \pm 0.01$, and moderate metallicity $[\mathrm{Fe}/\mathrm{H}] = -0.37\pm 0.03$.

Based on its ZTF light curve, we identify clear photometric variability with a well-defined period of $15.9665$ days (see folded light curve in Fig.~\ref{fig:258207200007316}). The absence of signatures indicating binarity or other types of variability suggests that the modulation is most likely due to rotational modulation from stellar spots. We therefore classify this object as a rotational variable (ROT), which with high probability belongs to the BY type.

\subsubsection{258208100007880}\label{star:258208100007880}

The star ZTF OID 258208100007880 (Gaia DR3 2912353727841154688, ATO J094.6486$-$24.8319) is classified as a close binary in the Atlas survey~\citep{2018AJ....156..241H} and as VAR by AAVSO VSX\footnote{\url{https://vsx.aavso.org/index.php?view=detail.top&oid=4468280}}. Based on ZTF DR23 data, we revised the period to $0.690844$ days and classified the source as EB. The updated folded light curve is shown in Fig.~\ref{fig:258208100007880}.

\subsubsection{258208100012201}\label{star:258208100012201}

The star ZTF OID 258208100012201 (Gaia DR3 2912282809341518336, ATO J094.9103-25.1659) is classified as  irregular variable star with $P=18.928$ days by \citet{2018AJ....156..241H} and long periodic variable by \citet{2022yCat.1358....0G}.

Based on this period and the shape of the ZTF folded light curve (Fig.~\ref{fig:258208100012201}), we suggest that the star is likely an ellipsoidal variable in a close binary system, which warrants a revision of its previous classification as an irregular variable.

\subsubsection{359205100018012}\label{star:359205100018012}

The star ZTF OID 359205100018012 (Gaia DR3 3046550595183360000, ZTF~J070713.15$-$103751.6) is classified as a semiregular variable by \citet{2020ApJS..249...18C}.

The ZTF light curve is shown in Fig.~\ref{fig:359205100018012}. Its blue color and the morphology of the light curve are consistent with a dwarf nova, yet the variability amplitude is significantly lower than typically observed for that class. 
Given its photometric properties, we also consider the possibility that it may be eruptive irregular variable of the $\gamma$ Cassiopeiae type (GCAS).

Interestingly, this object was also flagged as an anomaly by an Isolation Forest algorithm in the variability-based anomaly search by \citet{2022ApJ...932..118C}.

\subsubsection {359205200019829}\label{star:359205200019829}

\begin{figure*}[htbp]
    \centering

    \begin{subfigure}[b]{0.46\textwidth}
        \centering
        \includegraphics[width=\textwidth]{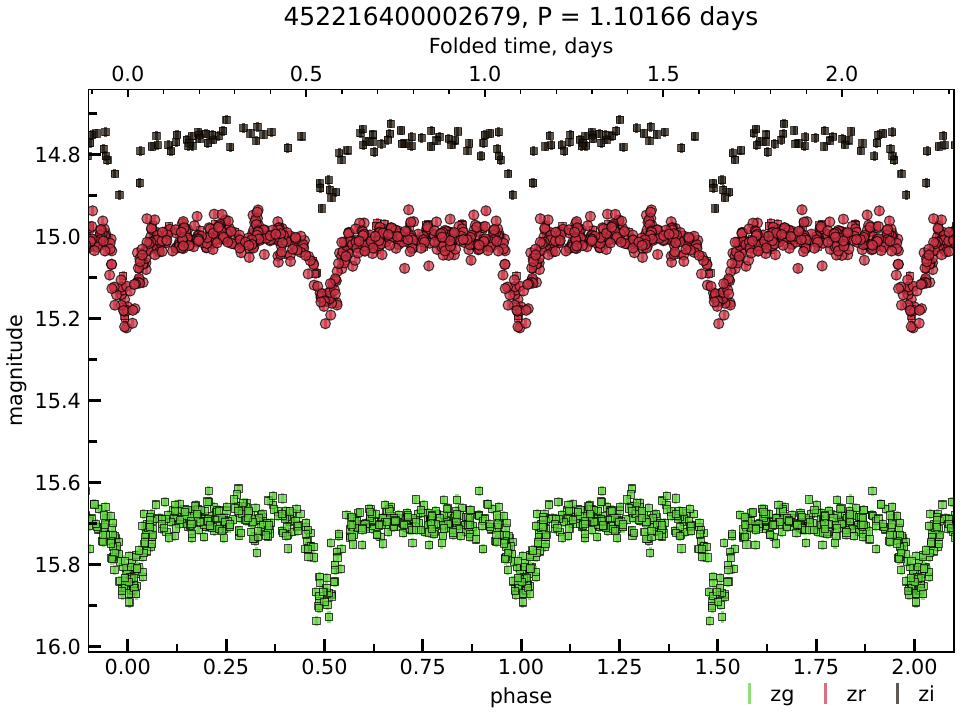}
        \caption{}
        \label{fig:452216400002679}
    \end{subfigure}
    \hfill
    \begin{subfigure}[b]{0.46\textwidth}
        \centering
        \includegraphics[width=\textwidth]{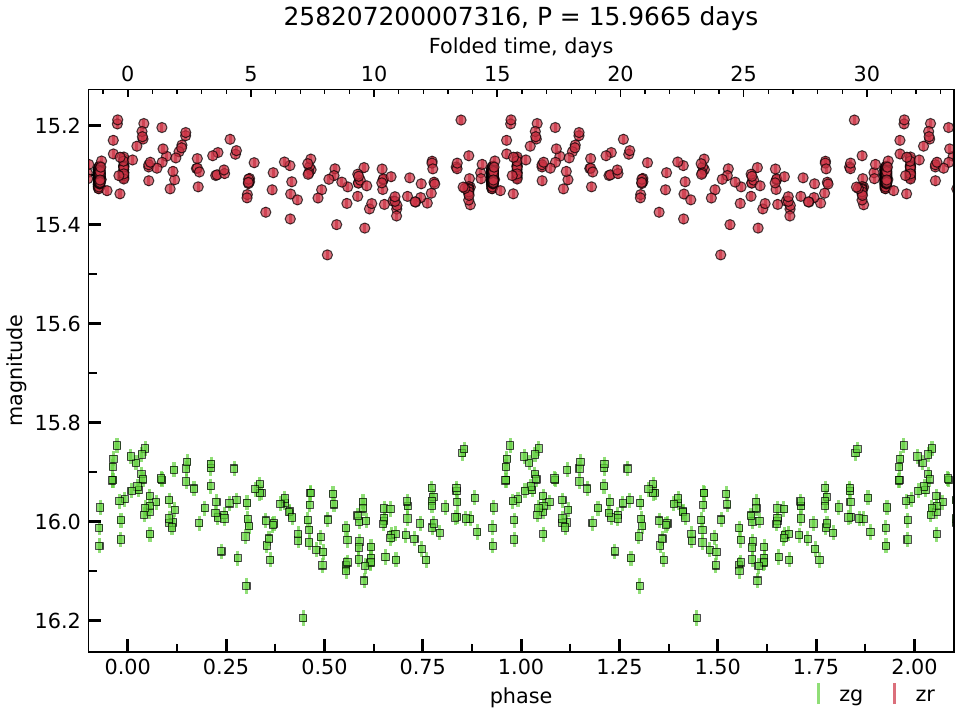}
        \caption{}
        \label{fig:258207200007316}
    \end{subfigure}

    \begin{subfigure}[b]{0.46\textwidth}
        \centering
        \includegraphics[width=\textwidth]{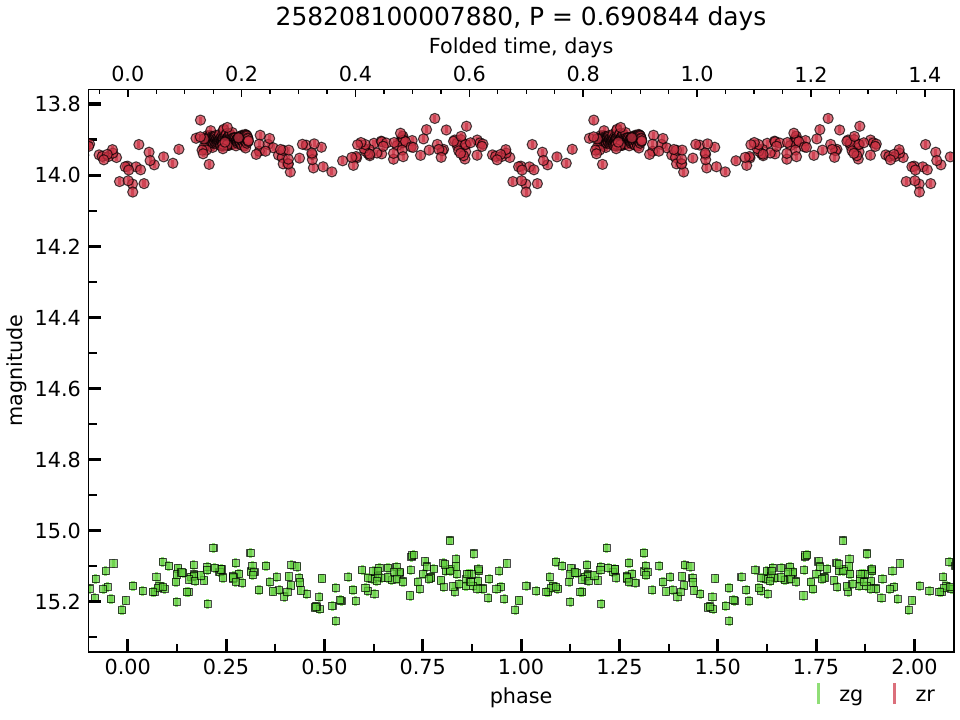}
        \caption{}
        \label{fig:258208100007880}
    \end{subfigure}
    \hfill
    \begin{subfigure}[b]{0.46\textwidth}
        \centering
        \includegraphics[width=\textwidth]{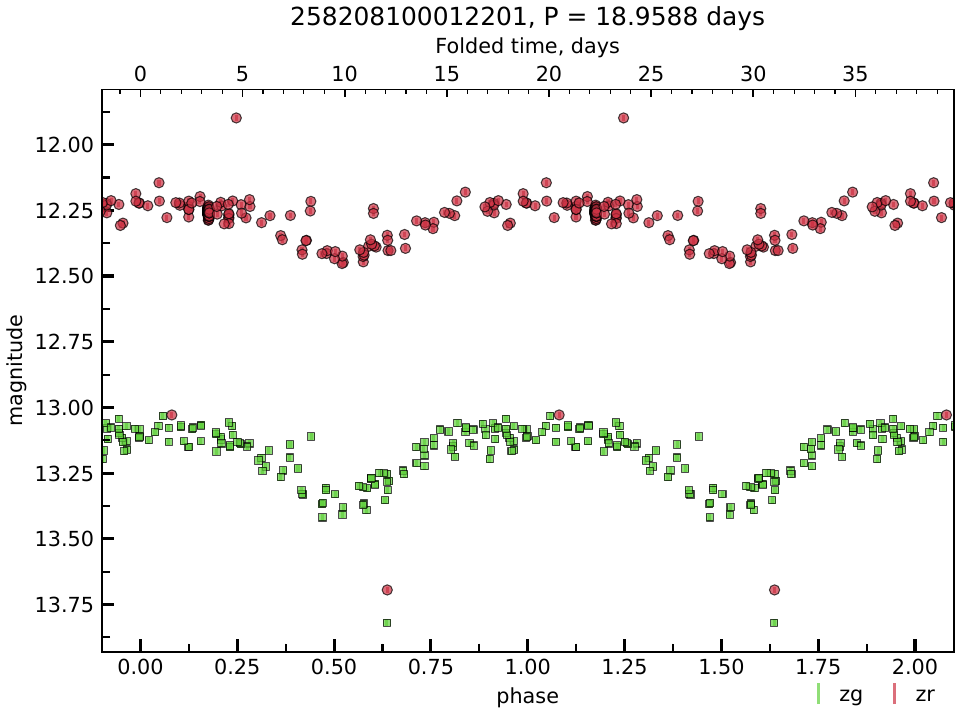}
        \caption{}
        \label{fig:258208100012201}
    \end{subfigure}

    \begin{subfigure}[b]{0.46\textwidth}
        \centering
        \includegraphics[width=\textwidth]{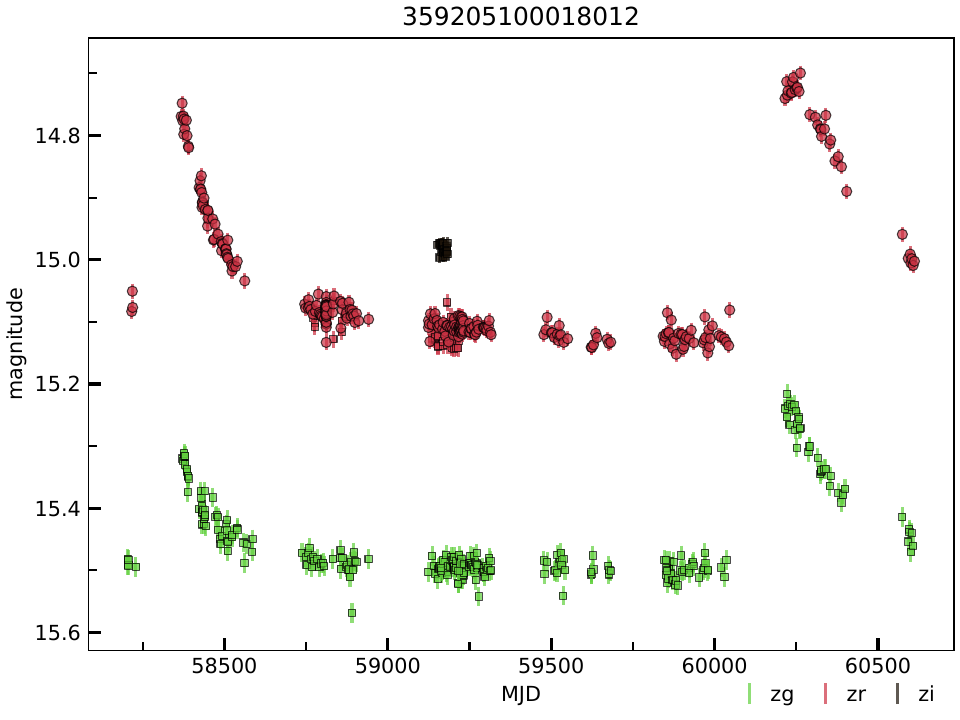}
        \caption{}
        \label{fig:359205100018012}
    \end{subfigure}
    \hfill
    \begin{subfigure}[b]{0.46\textwidth}
        \centering
        \includegraphics[width=\textwidth]{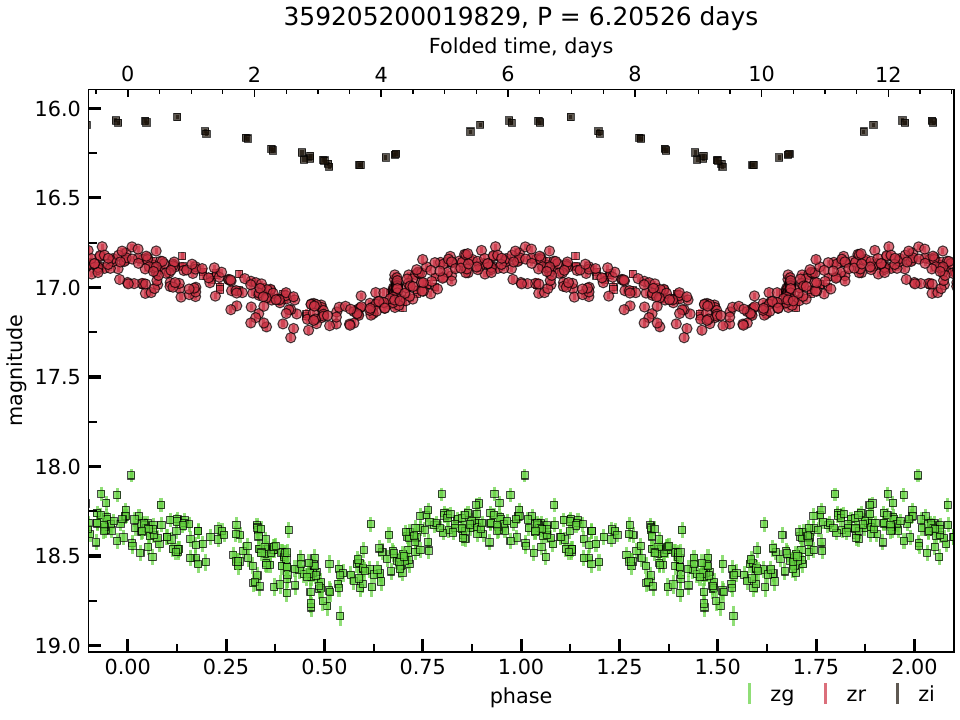}
        \caption{}
        \label{fig:359205200019829}
    \end{subfigure}

    \caption{Light curves of catalogued variables for which we refine the classification and/or determine the period:
    (\subref{fig:452216400002679}) 452216400002679 (see Section~\ref{star:452216400002679}),
    (\subref{fig:258207200007316}) 258207200007316 (see Section~\ref{star:258207200007316}),
    (\subref{fig:258208100007880}) 258208100007880 (see Section~\ref{star:258208100007880}),
    (\subref{fig:258208100012201}) 258208100012201 (see Section~\ref{star:258208100012201}),
    (\subref{fig:359205100018012}) 359205100018012 (see Section~\ref{star:359205100018012}),
    (\subref{fig:359205200019829}) 359205200019829 (see Section~\ref{star:359205200019829}).}
    \label{fig:combined}
\end{figure*}

The star ZTF OID 359205200019829 (GDS\_J0704458$-$104418, ATO~J106.1908$-$10.7383) is classified as ROT in \citet{2018MNRAS.477.3145J} and as a sinusoidal variable (NSINE) by \citet{2018AJ....156..241H}; however, no period was previously reported. We determined its period to be $6.20526$ days (Fig.~\ref{fig:359205200019829}). The smooth, low-amplitude, sinusoidal shape of the folded light curve supports its classification as a rotational variable. This interpretation is further supported by the physical parameters from Gaia: $T_\mathrm{eff} = 5485^{+84}_{-111}$~K, $\log g = 4.57 \pm 0.02$, and $[\mathrm{Fe}/\mathrm{H}] = -0.67^{+0.11}_{-0.10}$. Together, these characteristics are consistent with a BY variable, where the observed brightness changes are caused by star spots modulated by stellar rotation.

\section{Conclusions}
\label{sec:conclusions}

In this work, we presented the results of the SNAD-VIII Workshop, during which we applied the SNAD pipeline to search for anomalous light curves in the Zwicky Transient Facility Data Release 23, focusing on regions of the sky that overlap with the Vera C. Rubin Observatory's LSSTComCam target fields.

We ran the PineForest active anomaly detection algorithm on four LSSTComCam fields: two galactic (Rubin SV 95$-$25 and Seagull) and two extragalactic (Rubin SV 387 and ECDFS). In total, we visually inspected 400 ZTF light curves identified as anomalous by the algorithm (1 \% of the total number of objects).

As a result, we discovered six previously uncatalogued variable stars, including two showing solar-type variability, two RS~CVn candidates, one ellipsoidal variable, and one BY Draconis-type star. These objects have been added to the SNAD catalog and  assigned internal identifiers. In addition to the new discoveries, we also refined the classification and measured accurate periods for six known variable stars. 

Our findings show that in ZTF DR23, anomaly detection reveal new and interesting variables. With LSST’s deeper and richer data, we expect to discover many more rare and unusual objects using similar techniques.

\begin{acknowledgments}
This work is a result from the SNAD VIII Workshop\footnote{\url{https://snad.space/2025/}}, which took place in Yerevan, Armenia, from 29 June to 5 July 2025. We acknowledge the support of SberCloud\footnote{\url{https://cloud.ru/}} for providing computational resources for the development of this work, and in particular thank Albert Yefimov. We deeply thank Areg Mickaelian and all the staff of the Byurakan Astrophysical Observatory who kindly welcome our team during our stay in Yerevan. We thank Alexandra Zubareva for assistance with the classification of variable stars. M.~Pruzhinskaya and E.E.O. Ishida warmly thank Katyusha and Misha for their generous hospitality during our time in Armenia. M.~Pruzhinskaya, T.~Semenikhin, A.~Lavrukhina, M.~Kornilov acknowledge support from a Russian Science Foundation grant 24-22-00233, https://rscf.ru/en/project/24-22-00233/. The work of V.~Krushinsky was supported by a project of youth scientific laboratory, topic FEUZ-2025-0003. E.~E.~O.~Ishida  acknowledges support from the 2024 CNRS International Emerging Actions (IEA). Support was provided by Schmidt Sciences, LLC. for K.~Malanchev.
\end{acknowledgments}

\facilities{ZTF~\citep{https://doi.org/10.26131/irsa597}}

\software{Source Extractor \citep{1996A&AS..117..393B}, {\sc astropy}~\citep{2013A&A...558A..33A,2018AJ....156..123A,2022ApJ...935..167A}, {\sc NumPy}~\citep{numpy}, {\sc Matplotlib}~\citep{matplotlib}, {\sc pandas}~\citep{reback2020pandas,mckinney-proc-scipy-2010}.
}

\bibliography{biblio}{}
\bibliographystyle{aasjournalv7}

\end{document}